\documentclass{rmf-d}
\usepackage{nopageno,rmfbib,multicol,times,epsf,amsmath,amssymb,cite}
\usepackage[latin1]{inputenc}
\usepackage[]{caption2}
\usepackage{graphics}
\usepackage[demo]{graphicx}
\usepackage{hyperref}
\usepackage{lipsum}
\usepackage{wrapfig, blindtext}
\usepackage{comment}
\usepackage{tabularx}
\usepackage{float}
\def\rmfcornisa{Research OR Education of Physics \hfill\rmf\ {\bf ??} (*?*) ???--???
\hfill MES? A\~NO?}

\def\rmfcintilla{{\it Rev.\ Mex.\ Fis.\/} {\bf ??} (*?*) (????) ???--???}
\clearpage \rmfcaptionstyle 
\begin{document}
\title{Kaon and Nucleon States with Hidden Charm 
\vspace{-6pt}}
\author{Brenda B. Malabarba}
\address{Universidade de S\~ao Paulo, Instituto de F\'isica, C.P. 05389-970, Sao Paulo, Brazil}
\author{A. Mart\'{i}nez Torres}
\address{Universidade de S\~{a}o Paulo, Instituto de F\'{i}sica, C.P. 05389-970, S\~{a}o Paulo, Brazil}
\author{K.P. Khemchandani}
\address{Universidade Federal de S\~{a}o Paulo, C.P. 01302-907, S\~{a}o Paulo, Brazil}
\author{Xiu-Lei Ren}
\address{Institut fur Kernphysik \& Cluster of Excelence PRISMA$^+$, Johannes Gutenberg-Universitat Mainz, D-55099 Mainz, Germany}
\author{Li-Sheng Geng}
\address{ School of Physics, Beihang University, Beijing, 102206, China}
\maketitle
\recibido{day month year}{day month year
\vspace{-12pt}}
\begin{abstract}
\vspace{1em} In this talk we discuss the formation of exotic hadrons with hidden charm arising from three-body interactions. To be more specific, in the strangeness sector, we predict the existence of a mesonic state, $K^*(4307)$, which is dynamically generated from the three-body interactions of the $KD\bar{D}^*$ system, has mass around $4307$ MeV and quantum numbers $I(J^P) = 1/2\,(1^-)$. In the baryonic sector, we predict the existence of $N^*$ states, which are generated from the three-body interactions of the $ND\bar{D}^*$ system, with masses around $4400\sim 4600$ MeV, widths of $2\sim 20$ MeV and positive parity.

\vspace{1em}
\end{abstract}
\keys{Exotic States, Heavy-Quark Symmetry, Hidden Charm States \vspace{-4pt}}
\pacs{   \bf{\textit{14.20pt, 14.40pt}}    \vspace{-4pt}}
\begin{multicols}{2}

\section{Introduction}
In the latest years, due to the available access to higher energy regions, claims for the observation of exotic states have increased, drawing a lot of attention to the subject. A classical example of these exotic states are the so called $X$, $Y$ and $Z$ families (see, e.g., Refs. \cite{Hosaka:2016pey,Eberhard,Eulogio2016,stephen,richard}).

Interestingly, all heavy exotic hadrons found experimentally in the recent years have two characteristics in common: $1)$ the meson states can be interpreted as tetraquarks or as states obtained from the dynamics enclosed in two-meson systems, while the baryon states can be understood as pentaquarks or as states originated from meson-baryon systems $2)$ the mesons found have null strangeness.

In this talk we present the results obtained for two different systems with hidden charm which we have studied: $KD\bar D^*$ and $ND\bar D^*$. In the former case, such study was motivated by the fact that the $D\bar{D}^*$, $KD$, $KD^*$ subsystems have attractive interactions in s-wave, generating the states $X(3872)$ and $Z_c(3900)$ (in case of the $D\bar{D}^*$ interaction), $D_{s0}(2317)$ and $D_{s1}(2460)$ (from the $KD$ and $KD^*$ interactions). It is also interesting to note that considering all interaction in s-wave, the quantum numbers of the generated state are compatible with a $K^*$, but with a mass being in the charmonium sector. This fact is very interesting since, in spite of having every time more access to higher energy regions, there is a lack of data in the heavy strange mesonic sector (the last state reported by the PDG has a mass around $3100$ MeV \cite{pdg}). In case of the $ND\bar D^*$ system and the possible formation of $N^*$ states with hidden charm,  recently the LHCb collaboration announced the existence of possible hidden charm pentaquarks with non-zero strangeness and mass around $4459$ MeV \cite{mwang}. Such masses are close to the thresholds of  the $ND\bar{D}^*/N\bar{D}D^*$ system. As mentioned above, the interaction in the $D\bar{D}^*$ subsystem is attractive, generating the states $X(3782)$ and $Z_c(3900)$. Interestingly, the interactions in the $ND$ and $ND^*$ subsystems are also attractive and form, for example, the state $\Lambda_c(2595)$. In this way, it is quite probable to be able to generate $N^*$ states with a three-body nature as a consequence of the dynamics involved in the $ND\bar D^*$ system.

\section{Formalism}
We are interested in two different systems constituted by three hadrons. In order to study the dynamic of these system we can solve the Faddeev Equation to obtain the $T$-matrix for the system.

For a three-body system, if the third particle $P_3$ is lighter than the cluster composed by the two other particles ($P_1$ and $P_2$), and we are looking for the possible formation of bound states, we can rely on the fixed center approximation (FCA) to solve the Faddeev equations. 
Considering $N$ ($K$) as $P_3$ for the $ND\bar{D}^*$( $KD\bar{D}^*$) system, with $D\bar{D}^*$ clustering as $X(3872)$ or $Z_c(3900)$, with isospin 0 or 1 respectively, both our systems, $ND\bar{D}^*$ and $KD\bar{D}^*$, satisfies the above mentioned criteria, that is, for each of the systems $P_3$ is lighter than the cluster, such that we can use the FCA to solve the Faddeev Equations.

In the following, we are going to briefly present the formalism. More details can be found in Refs. \cite{malabarba1,malabarba2}.

Our goal is to obtain the three-body $T$-matrices for the $KD\bar{D}^*$ and $ND\bar{D}^*$ systems. By using the FCA we are able to decompose the amplitude $T$ as a sum of two partitions, $T_{31}$ and $T_{32}$, which satisfy the following coupled equations
\begin{align}
    T &= T_{31} + T_{32},\nonumber\\
    T_{31} &= t_{31} + t_{31}G_a T_{32},\nonumber\\
    T_{32} &= t_{32} + t_{32}G_a T_{31}.\label{fad}
\end{align}
In Eq.~(\ref{fad}), $t_{31}$ and $t_{32}$ are two-body $t$-matrices describing the interactions in the $KD(ND)$ and $K\bar{D}^*(N\bar{D}^*)$ subsystems, respectively, while $G_a$ is the propagator of the $P_3$ particle, that is, the propagator of $K\,(N)$, in the cluster, and it is given by
\begin{align}
    G_K &= \frac{1}{2M_a}\int \frac{d^3q}{(2\pi)^3}\frac{F_a(\mathbf{q})}{q_0^2 - \mathbf{q}^2 - m^2_K + i\epsilon},\nonumber\\
    G_N &= \frac{1}{2M_a}\int \frac{m_N}{\omega_N(\mathbf{q})}\frac{F_a(\mathbf{q})}{q_0 - \omega(\mathbf{q}) + i\epsilon}.\label{Gs}
\end{align}
In Eq.~(\ref{Gs}), $F_a$ is a form factor related to the molecular nature of the cluster and can be written as \cite{fator1,fator2,fator3} 
 \begin{eqnarray}
     F_a(\mathbf{q}) = \frac{1}{N}\int_{|\mathbf{p}|,|\mathbf{p - q}|<\Lambda} d^3\mathbf{p}\phantom{a}f_a(\mathbf{p})f_a(\mathbf{p -q}),\nonumber\\
     f_a(\mathbf{p}) = \frac{1}{\omega_{a1}(\mathbf{p})\omega_{a2}(\mathbf{p})}\cdot \frac{1}{M_a - \omega_{a1}(\mathbf{p}) - \omega_{a2}(\mathbf{p})},\label{FF}
 \end{eqnarray}
with $M_a$ being the mass of the cluster, $N = F_a(\mathbf{q} = 0)$ is a normalization constant, $\Lambda$ represents a cut-off $\sim$700 MeV and $\omega_{ai} = \sqrt{m_{ai}^2 + \mathbf{p}^2}$.

The only missing ingredients to solve Eq. \ref{fad} are the $t_{31}$ and $t_{32}$ two-body $t$-matrices. To illustrate the method for calculating them, we consider, for example, the $KD\bar D^*$ system and determine the expression for $t_{31}$.

The $KD\bar{D}^*$ system has three possible $K-$cluster isospin configurations: $\left|{KX,I=1/2,I_3=1/2}\right>$, $\left|KZ_c,I=1/2,I_3=1/2\right>$ and $\left|KZ_c,I=3/2,I_3=3/2\right>$. Considering for instance the state $\left|{KX,I=1/2,I_3=1/2}\right>$, the amplitude $t_{31}$ is given by 
\begin{align*}
\left<{KX,I=1/2,I_3=1/2}\right|t_{31}\left|{KX,I=1/2,I_3=1/2}\right>.
\end{align*}
Remembering that $t_{31}$ is related to the interaction between the $K$ and the $D$, it is useful to express the $KX$ state in terms of the isospin of the $KD$ subsystem. This can be done by using Clebsch-Gordan coefficients to write
\begin{align}
    &\left|{KX,I=1/2,I_3=1/2}\right> = \left|{KX,1/2,1/2}\right>\nonumber\\     
    &\quad=\frac{1}{2}\left[\left|KD, 1, 1\right>\otimes \left|\bar{D}^*,\frac{1}{2},-\frac{1}{2}\right>\right.\nonumber\\
    &\quad\left.-\frac{1}{\sqrt{2}}\left(\left|KD,1,0\right> + \left|KD,0, 0\right>\right)\otimes\left|\bar{D}^*,\frac{1}{2},\frac{1}{2}\right>\right]
\end{align}
Using the preceding equation, the $t_{31}$ amplitude for the $KX\to KX$ transition in isospin $1/2$ can be written as follows
\begin{equation}
    \left<{KX}\right|t_{31}\left|{KX}\right> \equiv {t_1}_{(11)} = \frac{1}{4}(3t_{KD}^{I = 1} + t_{KD}^{I = 0}),\label{KXstate}
\end{equation}
where, to simplify the notation, the subscript ${11}$ stands for $KX\to KX$ and $t_{KD}^{I = a}$ is the two-body $t$-matrix for the $KD$ subsystem with isospin $I = a$, where $a = 0,1$. The system $KZ_c$ can also have total isospin $1/2$, thus, it can couple to $KX$. This means that we also need to consider the transitions $KX\to KZ_c$ and $KZ_c\to KZ_c$ in order to obtain $t_{31}$. Repeating the process previously explained to get Eq.~(\ref{KXstate}), we obtain the results presented in Table \ref{resultados1}.
\begin{table}[H]
    \centering
    \begin{tabular}{c|c|c}
    &$KX$&$KZ_c$\\\hline
    $KX$&$\frac{1}{4}\left(3t_{KD}^{I=1} + t_{KD}^{I = 0}\right)$&$\frac{\sqrt{3}}{4}\left(t^{I = 1}_{KD} - t^{I = 0}_{KD}\right)$\\\hline
    $KZ_c$&$\frac{\sqrt{3}}{4}\left(t^{I = 1}_{KD} - t^{I = 0}_{KD}\right)$&$\frac{1}{4}\left(t_{KD}^{I = 1} + 3t_{KD}^{I = 0}\right)$ 
    \end{tabular}
    \caption{$t_{31}$ amplitudes for the $KD\bar{D}^*$ system in terms of the two-body $t$-matrices for the $KD$ subsystem.}
    \label{resultados1}
\end{table}
Analogously, in case of $t_{32}$ we obtain the same results of Table \ref{resultados1} but changing $D\to \bar{D}^*$ and adding a global minus sing on the non-diagonal terms.

Similarly, the results for the $ND\bar{D}^*$ system are completely analogous to those obtained for the $KD\bar{D}^*$ system once the same clusters are formed and both $K$ and $N$ have isospin $1/2$.

As we can see in Table~\ref{resultados1}, to determine $t_{31}$ and $t_{32}$, and thus solve Eq.~(\ref{fad}), we need the two-body $t$-matrices for the $ND/KD$, $N\bar{D}/K\bar{D}$, $ND^*/KD^*$, $N\bar{D}^*/K\bar{D}^*$ subsystems. These amplitudes can be obtained by solving the Bethe-Salpeter equation
\begin{equation}
    t_{AB} = V_{AB} + V_{AB}G_{AB}t_{AB},
\end{equation}
where $G_{AB}$ is the two body loop function for a channel made of hadrons $A$ and $B$ and $V_{AB}$ is the corresponding amplitude, which is obtained from an effective Lagrangian. The loop function $G_{AB}$ needs to be regularized either with cut-off or with dimensional regularization.

To determine the $KD$, $K\bar D^*$ two-body $t$-matrices, we have considered the following effective Lagrangian based on heavy-quark spin symmetry~\cite{Burdman:1992gh,Weinberg:1991um,Wise:1992hn}
\begin{equation}
    \mathcal{L} = \frac{1}{4f^2}\left\{\partial^\mu P\left[\phi,\partial_\mu P\right]P^{\dagger} - P\left[{\phi},{\partial_\mu}\right]\partial^\mu P^{\dagger}\right\},
\end{equation}
where $P$ and $\phi$ are given by
\begin{equation}
    P = \left(\begin{array}{ccc}
        D^0 & D^+ & D^+_s 
    \end{array}\right),
    \end{equation}
    \begin{equation}
    \phi = \left(\begin{array}{ccc}
    \frac{1}{\sqrt{2}}\pi^0 + \frac{1}{\sqrt{6}}\eta & \pi^+ & K^+ \\
    \pi^- & -\frac{1}{\sqrt{2}}\pi^0 + \frac{1}{\sqrt{6}}\eta& K^0\\
    K^- & \bar{K}^0 & -\frac{2}{\sqrt{6}}\eta
    \end{array}\right).
\end{equation}

In case of the $ND/N\bar{D}^*/ND^*/N\bar{D}$ two-body $t$-matrices, we consider two models: the first one is based on the $SU(4)$ and heavy-quark spin symmetries \cite{simsu41,simsu42}, while the second one is based on the $SU(8)$ spin flavor symmetry \cite{simsu81}.

It is interesting to note that different to the $KD$ and $KD^*$ systems, the $ND$ and $ND^*$ systems can be coupled: the state $ND$ ($ND^*$) on s-wave, i.e., orbital angular momentum $L=0$, has spin-parity $J^P = 1/2^-$ ($J^P = 1/2^-, 3/2^-$). Also, in d-wave, i.e., $L = 2$, the $ND$ ($ND^*$) system has $J^P = 3/2^-$ ($J^P = 1/2^-$). To consider this fact, in Ref.~\cite{simsu42} the transition amplitude $DN\to D^*N$ was extracted from box diagrams describing the processes $ND\to ND$ and $ND^*\to ND^*$ (for more details we refer the reader to Ref. \cite{simsu42}).

In the case of the model based on the $SU(8)$ spin flavor symmetry, the $DN\to D^*N$ transition is already included in the $SU(8)$ effective Lagrangian used in Ref.~\cite{simsu81}.

\section{Results}
\begin{figure}[H]
    \centering
    \includegraphics[width=.9\linewidth]{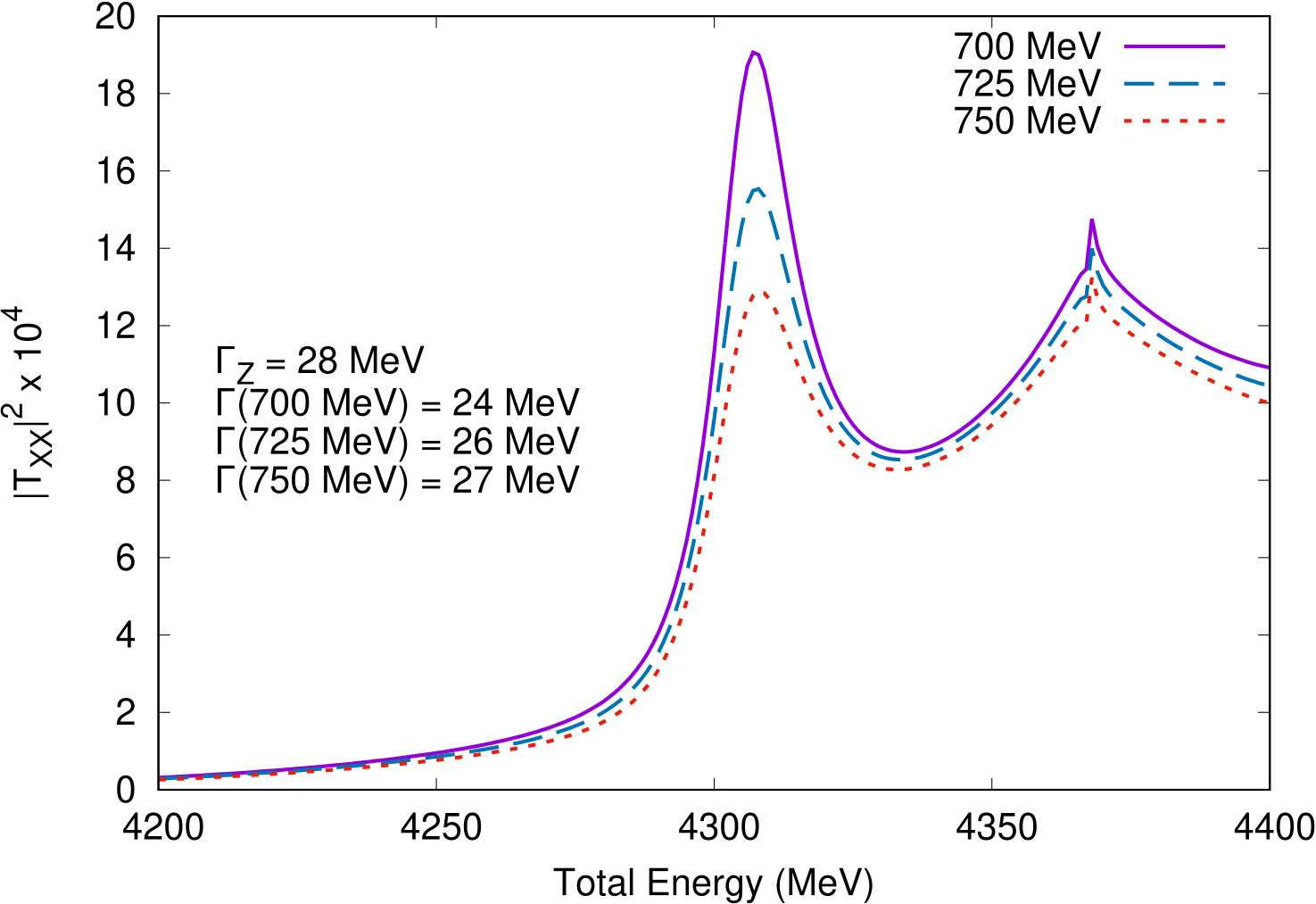}  
    \caption{Modulus squared of the three-body $T$-matrix for the transition $KX\to KX$ considering the coupling between the $KX$ and $KZ_c$ channels.}
\label{resultadoscomloopkdd1}
\end{figure}

\begin{figure}[H]
    \centering
    \includegraphics[width=.9\linewidth]{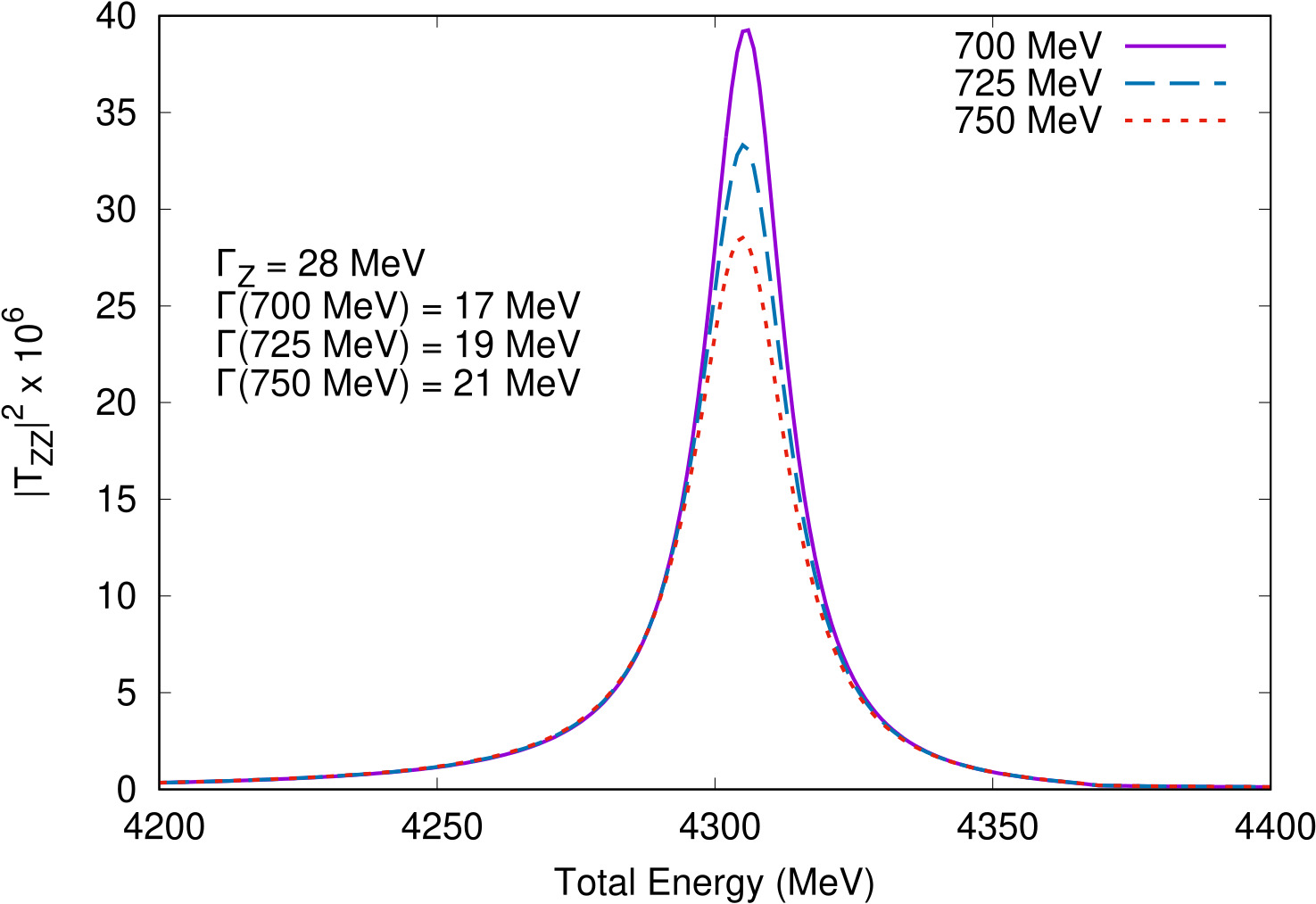}  
    \caption{Modulus squared of the three-body $T$-matrix for the transition $KZ_c\to KZ_c$ considering the coupling between the $KX$ and $KZ_c$ channels.}
\label{resultadoscomloopkdd2}
\end{figure}

In figures \ref{resultadoscomloopkdd1} and \ref{resultadoscomloopkdd2} we show the results obtained for $|T|^2$ as a function of $\sqrt{s}$ for the $KD\bar D^*$ system and the configurations $KX$ and $KZ_c$. As we can see, in both cases, a peak around $4300$ MeV shows up in the processes $KX\to KX$ and $KZ_c\to KZ_c$, considering $KX$ and $KZ_c$ as coupled channel when solving Eq.~(\ref{fad}). In the $KZ_c$ case, we have included the width of the $Z_c$ state by making the following transformation $M \to M - i\Gamma/2$, with $\Gamma\sim 28$ MeV, in the corresponding form factor of Eq.~(\ref{FF}). We have also varied the cut-off present in Eq.~(\ref{FF}) from $700$ to $750$ MeV, but as can be seen in the figures \ref{resultadoscomloopkdd1} and \ref{resultadoscomloopkdd2} there is not a strong dependence on the results with the cut-off.

In the $|T|^2$ of the $KX\to KX$ transition shown in Fig.~\ref{resultadoscomloopkdd1} we can also see a peak around $4375$ MeV, which is related to the threshold of the three-body system.

We have also done the computation of $|T|^2$ for the transition $KZ_c\to KZ_c$ with isospin $3/2$ but no clear signal of formation of a bound state was found.

\begin{figure}[H]
    \centering
    \includegraphics[width = 0.4\textwidth]{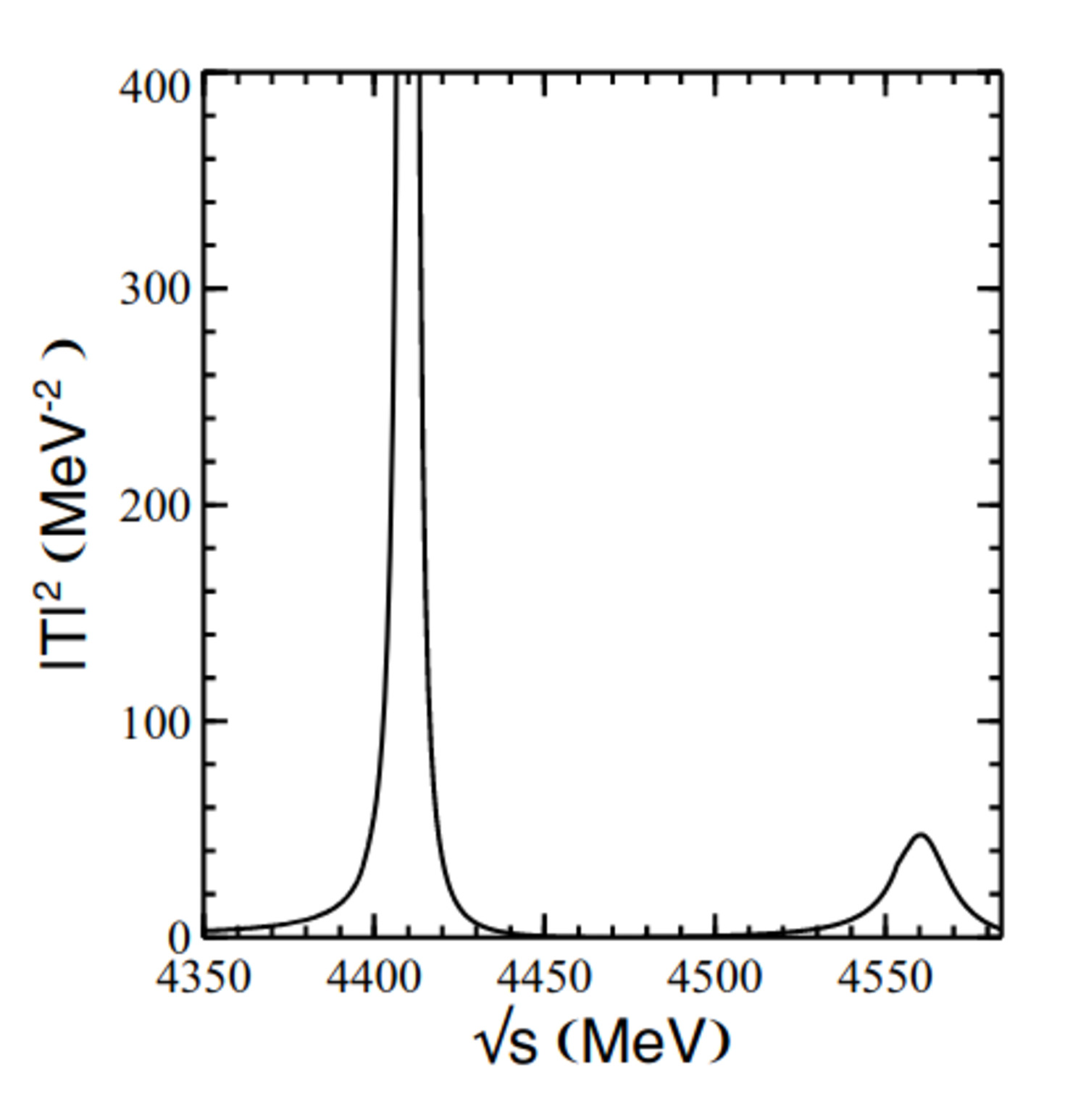}
    \caption{$|T|^2$ for the transition $NZ_c\to NZ_c$ with $I(J^P) = 1/2\,(1/2^+)$ as a functions of $\sqrt{s}$.}
    \label{resultadosNDD12b}
\end{figure}

\begin{figure}[H]
    \centering
    \includegraphics[width =0.4\textwidth]{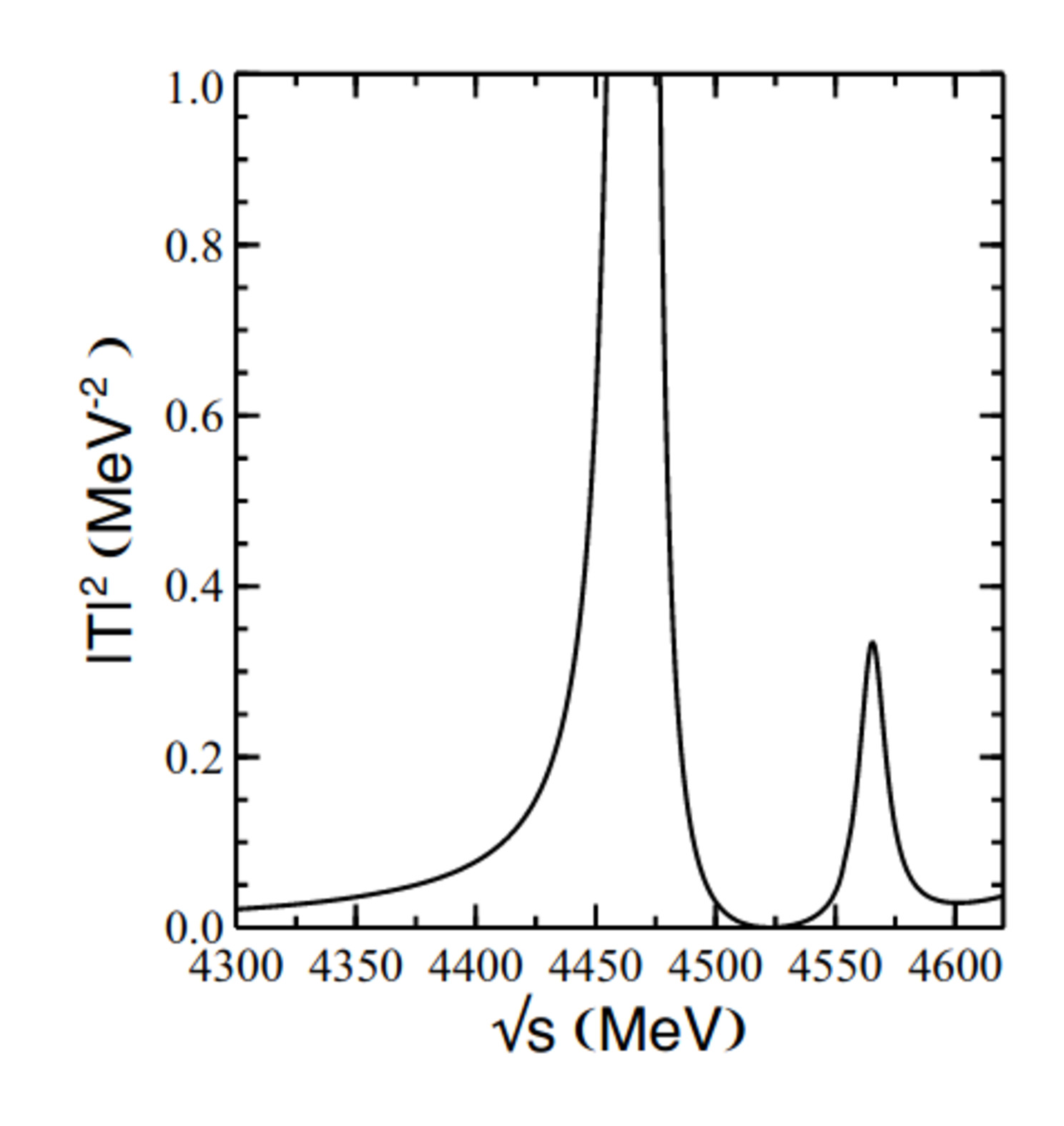}
    \caption{$|T|^2$ for the transition $NX\to NX$ with $I(J^P) = 1/2(3/2^+)$ as a functions of $\sqrt{s}$.}
    \label{resultadosNDD32a}
\end{figure}

\begin{figure}[H]
    \centering
    \includegraphics[width =0.4\textwidth]{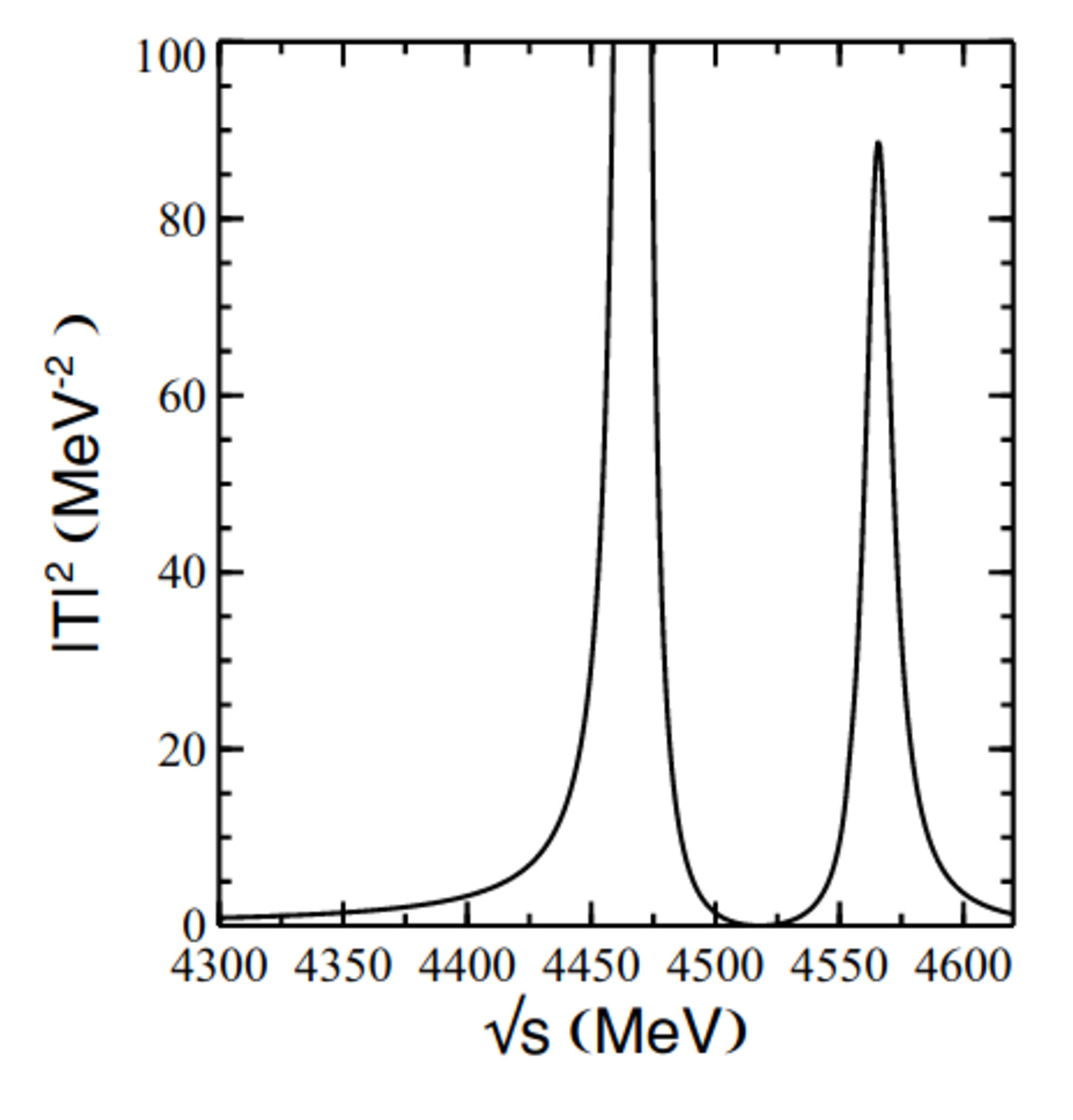}
    \caption{$|T|^2$ for the transition $NZ_c\to NZ_c$ with $I(J^P) = 1/2\,(3/2^+)$ as a functions of $\sqrt{s}$.}
    \label{resultadosNDD32b}
\end{figure}

\begin{figure}[H]
    \centering
    \includegraphics[width = 0.4\textwidth]{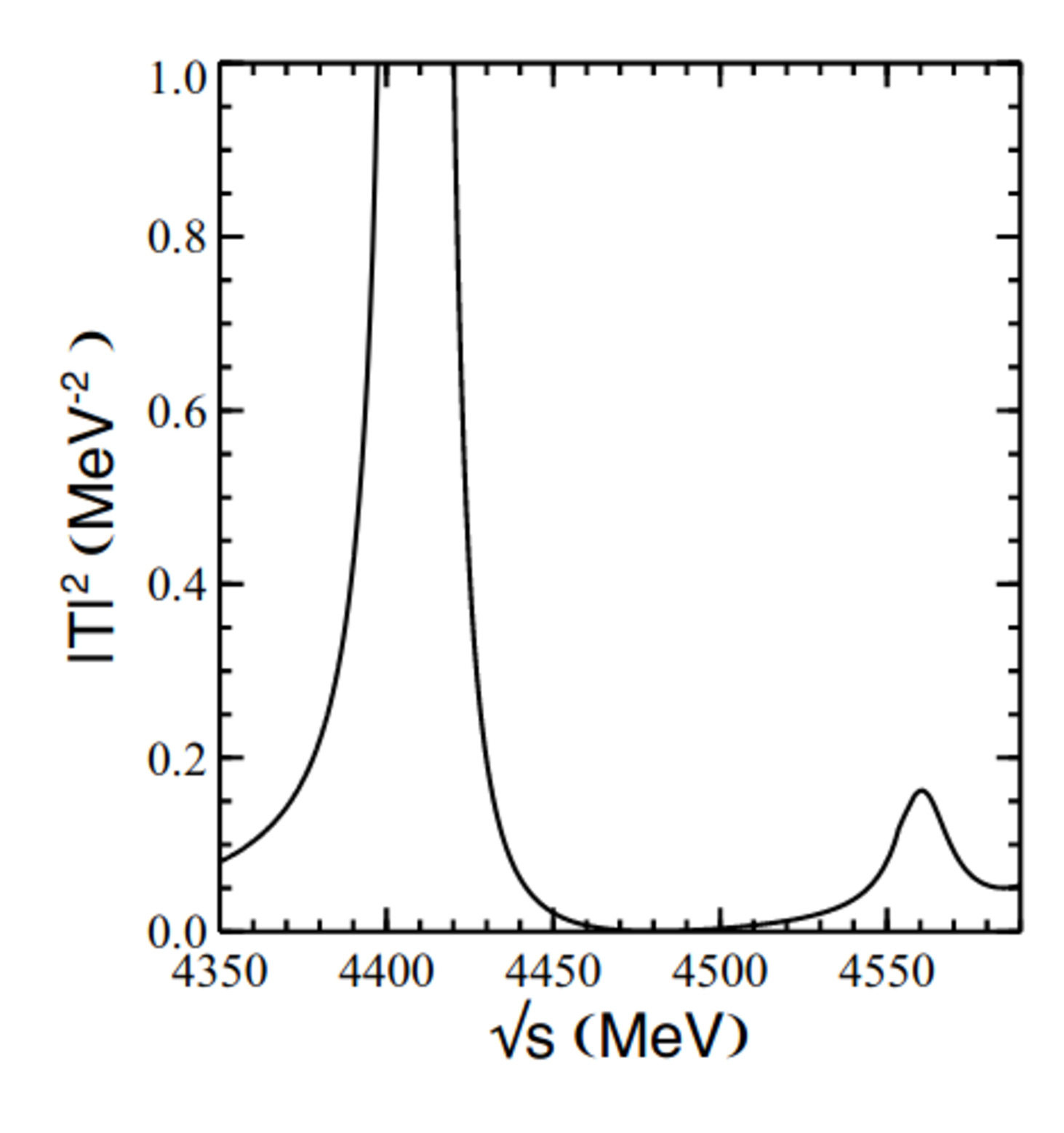}
    \caption{$|T|^2$ for the transition $NX\to NX$ with $I(J^P) = 1/2\,(1/2^+)$ as a functions of $\sqrt{s}$.}
    \label{resultadosNDD12a}
\end{figure}

In Figs.~\ref{resultadosNDD12b}$-$\ref{resultadosNDD12a}, we show the results obtained within the model of Refs.~\cite{simsu41,simsu42} for $|T|^2$ versus $\sqrt{s}$ for the $ND\bar{D}^*$ system. In case of Figs.~\ref{resultadosNDD12b} and \ref{resultadosNDD12a}, the results shown correspond to the transitions $NX\to NX$ and $NZ_c\to NZ_c$, respectively, with $I(J^P) = 1/2\,(1/2^-)$, while Figs. \ref{resultadosNDD32a} and  \ref{resultadosNDD32b} show the corresponding results in case of $I(J^P) = 1/2\,(3/2^+)$. In all cases $NX$ and $NZ_c$ are treated as coupled channels.
In all four graphics we can see that there are two peaks close to $4400$ MeV and $4550$ MeV, respectively. 
Varying the cut-off of Eq.~(\ref{FF}) from $700$ MeV to $770$ MeV causes a small shift, about $3-5$ MeV, on the masses of the states obtained. Similar results are found when using the model of Ref.~\cite{simsu81}. The results obtained with different models for the two-body interactions ($SU(4)$ heavy-quark spin symmetry model or $SU(8)$ model) as well as the different cut-off used in the form factor computation provide us uncertainties in the masses and widths of the states which we summarize in Table \ref{resultadossu4}.
\begin{table}[H]
    \centering
    \begin{tabular}{c|c|c}
        Spin-parity & Mass(MeV) & Width (MeV) \\
        $1/2^+$ & $4404 - 4410$ & $2$\\
        $1/2^+$ & $4556 - 4560$&$\sim 4-20$\\
        $3/2^+$ & $4467 - 4513$&$\sim 3-6$\\
        $3/2^+$ & $4558-4565$ & $\sim 5-14$
    \end{tabular}
    \caption{Masses and widths of the three-body states found in the study of the $ND\bar D^*$ system.}
    \label{resultadossu4}
\end{table}

\section{Conclusions and acknowledgments}
From our study we can conclude that adding a Kaon or a Nucleon to a cluster formed by $D\bar{D}^*$ generates states with hidden charm and a three-body molecular nature, that is, states whose inner structure can be described by the hadronic interactions without considering the quarks and gluons interactions.

Our findings for the $KD\bar{D}^*/K\bar{D}D^*$ system imply that a $K^*$ meson around $4307$ MeV should be observed in experimental investigation, 
while for the $ND\bar D^*$ system $N^*$ states with $J^P=1/2^+,3/2^+$ and masses around $4400-4600$ MeV are predicted.

This work has been supported by the
Funda\c{c}\~{a}o de Amparo \`{a} Pesquisa do Estado de S\~{a}o Paulo (FAPESP), processes number 2019/17149-3, 2019/16924-3 and 2020/00676-8, and by the Conselho Nacional Cient\'{i}fico e Tecnol\'{o}gico (CNPq), grant number 305526/2019-7 and 303945/2019-2.

\end{multicols}
\medline
\begin{multicols}{2}

\end{multicols}
\end{document}